\documentstyle[12pt]{article}
\begin{document}
{\Large\bf Masses of Quarks and Leptons and Mixing Angles in \\[2mm]
Anti-GUT Theory.}\\[2mm]
{\it C.D.~Froggatt$^a$ and H.B.~Nielsen$^b$}\\[2mm]
{\small $^a$University of Glasgow, $^b$Niels
Bohr Institute, Copenhagen}\\
\begin{abstract}
\noindent
We describe a fit to the charged fermion mass hierarchy using
the chiral quantum numbers of the maximal anti-grand unification
group $SMG^3 \times U(1)_f$, where $SMG \equiv
SU(3) \times SU(2) \times U(1)$. This fit suggests a set of
Higgs fields responsible for the breakdown, near the Planck
scale, of $SMG^3 \times U(1)_f$ to the Standard
Model group.
\end{abstract}

%\section{Introduction: The MPP and AGUT Model}
{\Large\bf 1 \ Introduction: The MPP and AGUT Model}\\[2mm]
Over many years we have gradually developed a model$^1$,
in which we are able to
understand or fit a large number of the parameters---coupling constants
and masses---in the Standard Model (SM), based on two assumptions which
we have called: Multiple Point Principle (MPP) and Anti Grand
Unified Theory (AGUT) respectively.  This model is not
specified in full detail, but allows, for example, a lot of
unspecified particles with masses of the order of the Planck scale.
There is a desert, with just SM interactions, essentially all the way
up to an order of magnitude or so under the Planck energy,
$M_{Planck}\simeq 10^{19}$ GeV. This is a larger energy range than most
physicists expect for the validity of the pure SM; in particular
we assume there is no supersymmetry in the desert.

The MPP can be formulated as the requirement
that there shall be many ``vacua'' with essentially the same energy
density; in the Euclideanised version of the theory, there
is a corresponding phase transition. This requirement of degenerate
vacua is then used to derive the values of various
coupling constants or relations between them. If this requirement
is imposed$^{2}$ on the pure SM with a cut-off close to
$M_{Planck}$, the values of the top quark and Higgs masses
($M_t$, $M_H$) must lie on the so-called vacuum stability curve.
In order for the vacuum degeneracy requirement to have a good
chance of being physically relevant, the vacuum expectation value (VEV)
of the Higgs field $\phi$ in the second vacuum (the VEV in the first
or usual vacuum is of course 246 GeV) must be of the same order
of magnitude as the cut-off.
This strongly first order phase transition condition
selects a particular point on the vacuum stability curve, giving our
SM predictions$^{2}$ for the top quark and Higgs boson pole
masses:
\begin{equation}
M_{t} = 173 \pm 5\ \mbox{GeV} \quad M_{H} = 135 \pm 9\ \mbox{GeV}
\end{equation}

The AGUT model is based on extending the SM gauge group,
$SMG = S(U(2) \times U(3)) \approx SU(3) \times SU(2) \times U(1)$,
in much the same way as grand unified $SU(5)$ but to the
non-simple gauge group $SMG^3 \times U(1)_f$, where the SM
gauge group is identified as the diagonal subgroup of the $SMG^3$
group. This means that, near the Planck scale, each of the three
quark-lepton generations has its own set of SM-like gauge particles
together with an additional abelian $U(1)_f$ gauge boson. The gauge
coupling constants are not unified but their values are
predicted$^{1}$ using the MPP principle. The $SMG^3$
gauge quantum numbers for the three quark-lepton generations are
assigned in the obvious way. We give their $U(1)_f$ charges, $Q_f$,
in section 2, where we discuss the AGUT group.
These new chiral gauge quantum numbers
distinguish between the three generations and we use their partial
conservation to naturally generate the charged fermion mass hierarchy.
The main rule of the game in our model is that any coupling
constant---at the fundamental level, the Planck scale presumably---is
of order unity, except for the Higgs field expectation values.
Therefore every quantity---such as effective running Yukawa couplings
at the Planck scale---is of order unity in the fundamental (Planck) units,
except for fermion mass suppression factors; these are
taken to be the product of the Higgs field VEVs,
counted in Planck units, over the number of Higgs fields
needed to provide the symmetry breaking (of our AGUT group
$SMG^3\times U(1)$) to make the mass matrix element in question non-zero.
Here, of course, any Higgs field that is needed several times delivers
its expectation value to the corresponding power. We even make
the assumption that every type of, say, fermion field needed with Planck
mass can be found: everything happens at the Planck scale, and with
unit strength! (This is contrary to some models in which the
lack of some types of, say, fermions at this scale plays an important role).
Roughly it is our philosophy that everything allowed can be found
at the Planck scale.
\\[2mm]

%\section{The Maximal AGUT Group} \label{sec:max}
{\Large\bf 2 \ The Maximal AGUT Group}
\\[2mm]
The $SMG^3\times U(1)_f$ group, with its 37 generators, at
first seems a rather arbitrary choice for a ``unified group''.
However it can be characterized uniquely as the gauge group $G$
beyond the SM containing the SM group and satisfying the
following 4 postulates:
\begin{enumerate}
\item $G \subseteq U(45)$. Here $U(45)$ is the group of all
unitary transformations of the 45 species of Weyl fields (3
generations with 15 in each) in the SM.
\item No anomalies. There should be neither gauge anomalies nor
mixed anomalies. We assume that only straightforward anomaly
cancellation takes place and, as in the SM itself, do not allow
for a Green-Schwarz type anomaly cancellation.
\item The various irreducible representations of Weyl fields
for the SM group remain irreducible under $G$. This
postulate is motivated by the observation that combining SM
irreducible representations into larger unified representations
introduces symmetry relations between Yukawa coupling constants,
whereas the particle spectrum exhibits a hierarchy between
essentially all the fermion masses rather than exact degeneracies.
\item $G$ is the maximal group satisfying the other 3 postulates.
\end{enumerate}

A rather complicated calculation shows that, modulo
permutations of the various SM fermion irreducible representations,
we are led to the result $G = SMG^3 \times U(1)_f$ with the usual
SM group embedded as the diagonal subgroup of $SMG^3$. Apart from
the various permutations of the particle names, the $U(1)_f$
group is unique. The $Q_f$ charges
can then be chosen so that the only non-zero values are carried
by the right-handed fermions of the second and third
proto-generations:
\begin{equation}
Q_f(\tau_R) = Q_f(b_R) = Q_f(c_R) = 1
\quad Q_f(\mu_R) = Q_f(d_R) = Q_f(t_R) = -1
\end{equation}

However we do have the freedom of choosing the gauge quantum numbers
of the Higgs fields reponsible for breaking the $SMG^3 \times U(1)_f$
group down to the SM group near the Planck scale. So we choose their
quantum numbers with a view to fitting
the fermion mass and mixing angle data,
extrapolated to the Planck scale using the SM renormalisation group
equations. We are thereby led to introduce three Higgs fields,
$W$, $T$ and $\xi$, with VEVs an order of magnitude or so below
$M_{Planck}$. In addition we introduce a Higgs field $S$ with a
VEV of order unity in Planck units. Furthermore we have to assign
AGUT gauge quantum numbers to the Weinberg-Salam Higgs field $\phi_{WS}$.
The existence of a field $S$, which does not suppress the fermion
masses, means that we cannot control phenomenologically when this
$S$-field is used in the mass matrices.
Thus all the quantum numbers of the other Higgs fields, found
by fitting data, can only have their quantum numbers predicted
modulo those of the field $S$. We specify the Higgs field abelian
quantum numbers as a charge vector $\vec{Q} \equiv \left ( y_1/2,
y_2/2,y_3/2,Q_f \right )$, where $y_i/2$ denotes the weak
hypercharge for the $i$'th proto-generation. We then determine their
non-abelian representations, by imposing the natural generalisation
of the SM charge quantisation rule:
\begin{equation}
y_i/2 + d_i /2 + t_i/3 = 0 \quad ( \mbox{mod} \quad 1)
\label{quantrule}
\end{equation}
We also require that the non-abelian representations be the smallest
possible (singlet or fundamental like the fermions)
with the dualities $d_i$ and/or trialities $t_i$ determined
from the quantisation rule of eq.~(\ref{quantrule}).
\\[2mm]

%\section{Fermion Masses and Mixing Angles}
{\Large\bf 3 \ Fermion Masses and Mixing Angles}\\[2mm]
We have chosen the Higgs field quantum numbers up to the above-mentioned
ambiguity modulo those of the field $S$ and obtained$^{3,4}$
the following order of magnitude effective SM Yukawa coupling matrices:
\begin{equation}
\label{YUD}
Y_U \simeq \pmatrix{WT^2\xi^2 & WT^2\xi & W^2T\xi \cr
				   WT^2\xi^3 & WT^2    & W^2T    \cr
				   \xi^3     & 1       & WT \cr}
\qquad
Y_D \simeq \pmatrix{WT^2\xi^2 & WT^2\xi & T^3\xi  \cr
				   WT^2\xi   & WT^2    & T^3     \cr
				   W^2T^4\xi & W^2T^4  & WT	\cr}
\end{equation}
for the up and down type quarks, and for the charged leptons we have:
\begin{equation}
\label{YE}
Y_E \simeq \pmatrix{WT^2\xi^2 & WT^2\xi^3 & WT^4\xi \cr
				   WT^2\xi^5 & WT^2    & WT^4\xi^2 \cr
				   WT^5\xi^3 & W^2T^4  & WT \cr}
\end{equation}
Here $W$, $T$ and $\xi$ denote the VEVs of the Higgs field in
Planck units. By including the order of one $S$ field VEV in
the fit\footnote[2]{We also included a factorial factor in
each matrix element keeping track of the number of permutations of the
Higgs fields mediating the corresponding quantum number transition;
these factorials essentially have the effect of renormalising the
VEVs in the fit.}
we get a set of results dependent on the ambiguity in
the quantum number choice. In table \ref{bestfit}, we present the
results for the following set of quantum numbers, chosen on the
principle of using small representations:
\begin{displaymath}
\vec{Q}_{\phi_{WS}} = (1/6 ,1/2 ,-1/6,0) \quad
\vec{Q}_W = (-1/6 ,-1/3,1/2,-1/3) \quad
\vec{Q}_T = (-1/6,0,1/6 , 1/3)
\end{displaymath}
\begin{equation}
\vec{Q}_{\xi} = (0,0,0,1) \quad
\vec{Q}_S = ( 1/6,-1/6, 0 ,-1)
\end{equation}
The quantum numbers $\vec{Q}_{\phi_{WS}}$ have been chosen to ensure
that the order of one top quark Yukawa coupling
corresponds to an off-diagonal element of $Y_U$.
\begin{table}[t]
\caption{Best fit to experimental data. All masses are running masses at 1 GeV
except the top quark mass which is the pole mass.}
\begin{displaymath}
\begin{array}{|c|c|c|c|c|c|c|}
\hline
 & m_u & m_d & m_e & m_c & m_s & m_{\mu}  \\ \hline
{\rm Fitted} & 2.9 {\rm \; MeV} & 9.9 {\rm \; MeV} &
0.71 {\rm \; MeV} & 0.98 {\rm \; GeV} & 426 {\rm \; MeV} &
88 {\rm \; MeV} \\ \hline
{\rm Experimental} & 4 {\rm \; MeV} & 9 {\rm \; MeV} &
0.5 {\rm \; MeV} & 1.4 {\rm \; GeV} & 200 {\rm \; MeV} &
105 {\rm \; MeV}  \\ \hline
\end{array}
\end{displaymath}
\begin{displaymath}
\begin{array}{|c|c|c|c|c|c|c|}
\hline
 & M_t & m_b & m_{\tau} & V_{us} & V_{cb} & V_{ub} \\ \hline
{\rm Fitted}   & 153 {\rm \; GeV} & 7.4 {\rm \; GeV} &
1.35 {\rm \; GeV} & 0.16 & 0.030 & 0.0033 \\ \hline
{\rm Experimental}  & 180 {\rm \; GeV} & 6.3 {\rm \; GeV} &
1.78 {\rm \; GeV} & 0.22 & 0.041 & 0.0035 \\ \hline
\end{array}
\end{displaymath}
\label{bestfit}
\end{table}

The most characteristic feature of the AGUT Yukawa matrices $Y_U$,
$Y_D$ and $Y_E$ is that their diagonals are equal order of
magnitudewise. This feature follows from the quark-lepton
quantum numbers, which all follow from the general structure of
the model, and is independent of the choice of Higgs fields.
Apart from the top and charm quarks, the fermion mass eigenvalues
are given in order of magnitude by the diagonal elements and
hence the AGUT model simulates the GUT SU(5) mass predictions,
namely the degeneracy of the $dsb$-quarks with the charged leptons
in the corresponding generations. Note, however, that we
only get the prediction of these degeneracies at the Planck scale
as far as order of magnitude is concerned, and not exactly!
This gives much better agreement with experiment than exact
SU(5) predictions, which are rather bad unless more
Weinberg-Salam Higgs fields are included
a la Georgi-Jarlskog's factor 3 mechanism.
Also note that we in addition predict that the up-quark
is degenerate with the down-quark and the electron.
This does not follow
just from GUT SU(5), although the up-quark
is equally, not to say better,
degenerate with the electron than the down quark!

It is possible to obtain rather simple relations from our model by
eliminating the suppression factors. First one gets the already
mentioned degeneracy of the masses in the same generation, except
for the top and the charm quarks (all after transport
by the renormalisation group to the Planck scale).
In addition we have the following order
of magnitude Planck scale relations:
\begin{eqnarray}
m_b^3 \simeq m_t m_c m_s \qquad \qquad
V_{ub} \simeq  V_{td} \simeq V_{us} V_{cb}\\
V_{us} \simeq V_{cd} \simeq \sqrt{ \frac{ m_d}{m_s}} \qquad \qquad
V_{cb} \simeq V_{ts} \simeq \frac{ m_s^2}{ m_c m_b }
\end{eqnarray}
We also predict the CP-violating area of the ``unitarity triangle''
to be given order of magnitudewise by $J \simeq V_{us} V_{cb} V_{ub}$.
\\[2mm]

{\Large\bf References}\\[2mm]
{\small
1.\ D.L. Bennett, C.D. Froggatt and H.B. Nielsen, hep-ph/9710407;
To be published in the Proceedings of the APCTP-ICTP Joint
International Conference (AIJIC 97) on Recent Developments in
Nonperturbative Quantum Field Theory, Seoul, Korea,
26-30 May 1997.\\
2.\ C.D. Froggatt and H.B. Nielsen,
Phys. Lett. {\bf B 368} (1996) 96.\\
3.\ C.D. Froggatt, H.B. Nielsen and D.J. Smith,
Phys. Lett. {\bf B 385} (1996) 150.\\
4.\ C.D. Froggatt, M. Gibson, H.B. Nielsen and D.J. Smith,
hep-ph/9706212.\\
}
\normalsize
\end{document}